\documentclass[iop]{emulateapj-rtx4} 
\shortauthors{Sekanina}
\shorttitle{Odd Fragmented Comet 157P}
\slugcomment{Version \today }

\begin{document}
\title{The Odd Comet 157P/Tritton and Its Misunderstood Fragmentation\\[-1.6cm]}
\author{Zdenek Sekanina}
\affil{La Canada Flintridge, California 91011, U.S.A.; {\it ZdenSek@gmail.com}} 

\begin{abstract} 
Comet 157P is a faint object with a history of being prone to unfortunate
situations, circumstances, and/or coincidences.  Several weeks after its 1978
discovery the comet disappeared and remained lost nonstop for twenty five
years.  Rediscovered in 2003 as a new comet, it was about 500 times brighter
than in 1978,  caught apparently in one of its outbursts. The comet was not
detected 200 days after its 2016 perihelion, being fainter than mag~20, but
80~days later it was mag~16 and gradually fading back to mag~20 over a period of
four months.  The comet did not miss the opportunity to have a close encounter
with Jupiter, having approached it to less than 0.3 AU on 2020 February 10.~The
2017 outburst or surge of activity appears to have accompanied an event of
nuclear fragmentation.  The birth of a second companion is dated to the months
following the Jupiter encounter. The series of weird episodes culminated near
the 2022 perihelion, when one companion brightened to become observable for
two weeks and after another two weeks the other flared up to be seen for the
next two weeks.  Unnoticed, this incredible coincidence fooled some experts
into believing that~a~single~object,\,designated~\mbox{157P-B}, was involved,
even though its orbit left large residuals.  I now offer representative
fragmentation~solutions for the two companions, the mean residuals amounting to
$\pm$0$^{\prime\prime}\!$.4 and $\pm$1$^{\prime\prime}\!$.0, respectively.
\end{abstract}
\section{A Short, Peculiar History of Comet Tritton} 
%
Even though periodic comet Tritton, officially designated 157P, was
discovered only 45~years ago, it appears to have experienced strange
events and have become part of bizarre situations at a rate higher
than any other periodic comet.  The single exception is that it has
not~disintegrated, not yet.

The unusual incidents began with the discovery plate taken by K.\ Tritton,
U.K.\ Schmidt Telescope~Unit,~Coonabarabran,\,on 1978 February 11:\
Marsden\,\&\,Green\,(1985) remarked that this was the first time ever that
images of {\it three\/} comets were known to have appeared on the same
exposure.  Designated as comet 1978d, Tritton shared~the plate with
4P/Faye and C/1977~D1 (Lovas).

Since Tritton was faint, of apparent magnitude~\mbox{19--20}, and, as it
turned out, some 3.5~months after perihelion at discovery, few observations
were made in 1978 and a total of only seven astrometric positions became
available, spanning an arc of 31~days.  And even though the comet received
the definitive designation 1977~XIII (Marsden 1979), its orbital period
was not determined accurately enough to provide a reliable ephemeris for
the next peri\-helion return in 1984.

The situation was in fact much more critical, because the comet was
missed not only in 1984, but in 1990 and 1996 as well.  It became
a long-lost comet and in the new designation system, introduced in
the mid-1990s, Tritton was referred to as comet D/1978~C2 (e.g.,
Marsden~1995), where the prefix D meant that the object was unworthy
of serious recovery efforts.

In 2003, P.\ Holvorcem, Campinas, Brazil, reported~the discovery of
a fast-moving object by C.\ W.\ Juels, Fountain Hills, Arizona, with
a 12-cm refractor on~\mbox{October 6}.  For a day or so, before
identity was established (Green 2003), the accidentally rediscovered
comet Tritton had been masquerading as comet Juels or Juels-Holvorcem.
The prefix D was dropped and, yet again, the comet was assigned
a new designation --- P/2003~T1.  Compared to 1978, the comet was
substantially brighter at this apparition, reaching a peak magnitude
of approximately 11, even though it was nearly 1.7~AU from the Earth
when first detected.

The rediscovery was instrumental in that the orbital period could
accurately be determined and the comet prevented from getting lost
again.  Indeed, since 2003~it has been observed at every return.
Monitored extensively, it did not get brighter than magnitude $\sim$15
near perihelion in 2010.  Worse yet, it was of magnitude~20 when
recovered more than nine months before its perihelion in 2016.
Interesting developments took place following the perihelion passage:\
whereas H.\ Sato failed to detect the comet on 2016 December~28, about
200~days after perihelion, on an exposure that reached
magnitude~20,\footnote{See\,website {\tt
https://groups.io/g/comets-ml/messages},\,Sato's message \#26432
dated 2017 April~24.} the object was near magnitude~16 and getting
rapidly fainter some 80~days later, when it already was more than 3~AU
from the Sun.

In its most recent display of bravado, Tritton passed 0.265~AU from
Jupiter on 2020 February~10,\footnote{See \,website \,{\tt
https://ssd.jpl.nasa.gov/tools/sbdb\_lookup.}{\linebreak}
{\tt html\#/?des=157P}.} and upon arrival at its September 2022
perihelion it exhibited a secondary nucleus between August~21
and September~2 and again between September~15 and 28.  As I show in
Section~3, suspicion based on cursory data inspection that the comet
was preparing yet another surprise for us is supported by computations.

Examination of the physical behavior of comet Tritton since Sato's
nondetection in late 2016 and presentation of a hypothesis for the
reported features are the objectives of this paper.

\section{Post-Perihelion Light Curve in 2016 Return} 
The observing conditions during the 2016 return of comet Tritton
were singularly unfavorable.  The perihelion occurred on June 10,
and between late February and late October the comet stayed at
solar elongations smaller than 30$^\circ$.  Observations made long
before perihelion, in September--December 2015 showed the comet at
nuclear magnitude~$\sim$20 with only minor variations.

The noted negative observation by Sato in late December 2016 suggested
that at heliocentric distances near 2.4~AU the comet was even fainter
after perihelion that before it.  But when the observers at the
ATLAS-HKO Station on Haleakala, Maui, pointed their 50-cm f/2 Schmidt
at the comet on 2017 March~19.6~UT, about 80~days after Sato, it was
brighter than magnitude 17.  From March~23 on the comet was observed
worldwide; the astrometric and brightness data have been published by
the Minor Planet Center (MPC Staff 2017).

\begin{figure}[b]
\vspace{1.7cm}
\hspace{-0.54cm}
\centerline{
\scalebox{1.315}{
\includegraphics{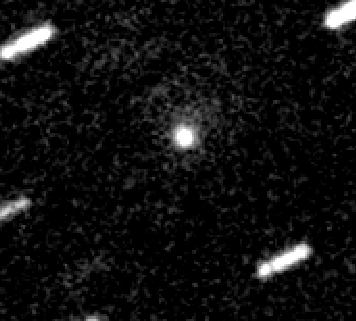}}} 
\vspace{0cm}
\caption{Comet 157P/Tritton imaged by J.-F.\ Soulier with his 30-cm
f/3.8 reflector at Maisoncelles, France, on 2017 April 27.96 UT.
Note the disk-like appearance, with only a faint coma.  The field
measures 5$^{\prime}\!$.5 along the diagonal.  North is up, east to the
left. (Courtesy of J.-F.\ Soulier.){\vspace{0cm}}}
\end{figure}

The unusual feature of these data is that, unlike for other
comets, most observers have reported the total, rather than nuclear,
magnitudes.  The anomaly is readily understood when one inspects the
comet's appearance at the time.  In Figure~1 I reproduce an image
obtained by J.-F.\ Soulier with his 30-cm f/3.8 reflector at
Maisoncelles, France, on April~27.  The picture shows the comet as
a disk-like object approximately 15$^{\prime\prime}$ across with
only traces of a very faint coma.

\begin{table}[t]
\vspace{0.17cm}
\hspace{-0.2cm}
\centerline{
\scalebox{1}{
\includegraphics{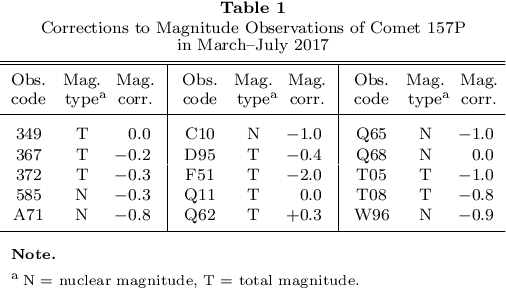}}} 
\vspace{0.7cm}
\end{table}

The comet was observed from March~19 until July~18 from about two
dozen observatories, from 15 of them more than once.  (By a single
observation is meant here an average magnitude from a set of two,
three or more exposures separated from one another by tens of seconds
to several minutes, as is common practice these days.)  Since the
aperture sizes and color systems of the measured magnitudes differ,
the reported results by different observers vary.  To reduce the
scatter, I refer the magnitude observations from the 15 observatories
to a common system by applying empirical corrections.  A reported
magnitude $H({\sf code,type})$, where the type is either a nuclear
(N) or total (T) magnitude, is thus converted to a standard apparent
magnitude $H_{\rm app}$ by
\begin{equation}
H_{\rm app} = H({\sf code,type}) + {\sf corr.(code,type).} 
\end{equation}
The applied corrections are presented in Table 1, where I chose ${\sf
corr.} \!=\! 0$ for the magnitudes reported by K.\ Kadota ({\sf code} 349).
From my earlier experience, his CCD magnitudes have fairly consistently
been about 1 to 1.5~mag fainter than visual estimates made with the
naked eye or small binoculars.

The 48 data points of $H_{\rm app}$ obtained in this fashion are
plotted as a function of time in Figure~2.  For comparison, Sato's
nondetection is also depicted, showing that the comet's outburst
or flare-up  had an amplitude of at least 4~mag and that it began
between early January and mid-March of 2017.  From mid-April on, the
brightness was subsiding at an average rate of 0.043~mag per day;
thus, the fading was rapid but not precipitous.

\begin{figure}[b]
\vspace{0.93cm}
\hspace{-0.21cm}
\centerline{
\scalebox{0.655}{
\includegraphics{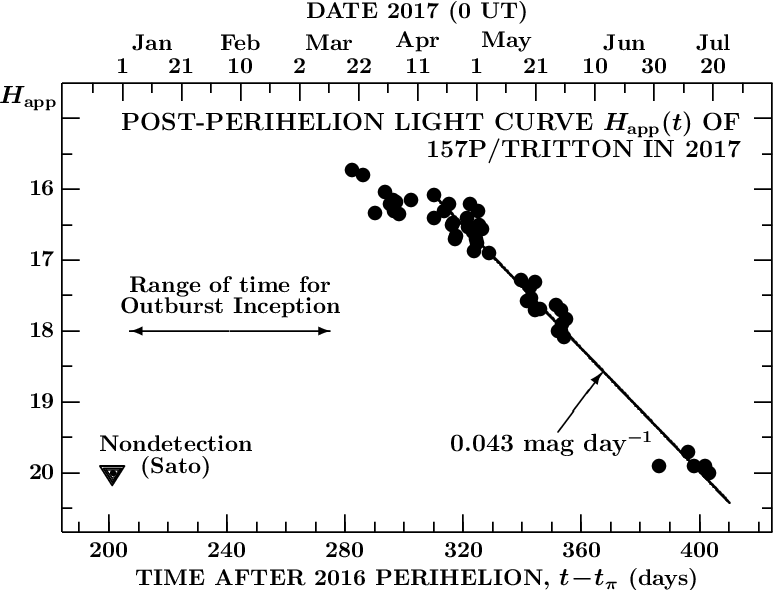}}} 
\vspace{-0.1cm}
\caption{Post-perihelion light curve of 157P based on 48 observations
made between 2017 March~19 and July~18.  Comparison with Sato's
nondetection in late December 2016 suggests an amplitude of the outburst
of at least 4~mag.  After mid-April the comet was fading rapidly, at
an average rate of 0.043 mag per day.{\vspace{0cm}}}
\end{figure}

Figure 3 shows a plot of the corrected magnitude, now normalized to
a distance of 1~AU from the Earth, $H_{\!{_\Delta}}$, as a function of
heliocentric distance, $r$ (on a log scale).  Resembling Figure~2,
it demonstrates that the outburst began between 2.4~AU and 2.9~AU
from the Sun, probably closer to the latter limit.  The fading after
mid-April followed on the average an inverse 17th power of heliocentric
distance, confirming that it was very steep but~not abrupt.

\begin{figure}[b]
\vspace{0.97cm}
\hspace{-0.19cm}
\centerline{
\scalebox{0.69}{
\includegraphics{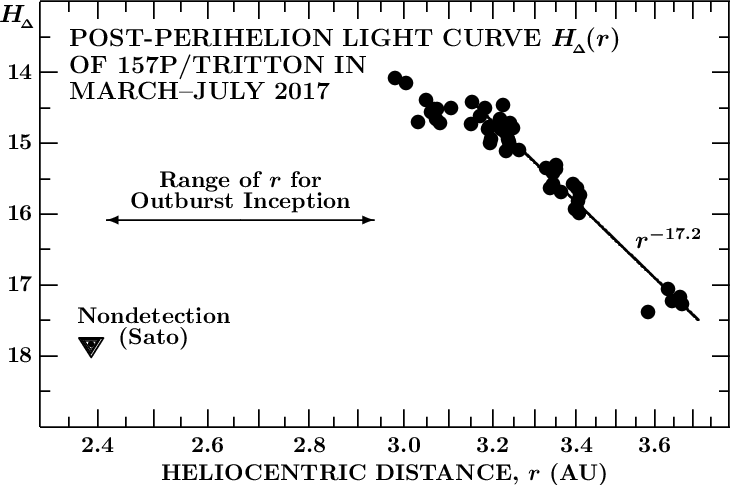}}} 
\vspace{-0.1cm}
\caption{Post-perihelion light curve of comet 157P, based on the
48~observations made between 2017 March~19 and July~18.  The
magnitude has been normalized to 1~AU from the Earth.  Comparison
with Sato's nondetection suggests that the outburst began between
2.4~AU and 2.9~AU from the Sun.  Note that beyond 3.1~AU the
comet's elevated normalized brightness subsided approximately
as an inverse 17th power of heliocentric distance.}
\end{figure}

Comet Tritton was observed on a number of occasions in 2021, between
July~17 and October~26, 419 to 318~days before the 2022 perihelion,
always of magnitude 20--21 (MPC Staff 2021, 2022a).  It was next
detected on 2022 July~12, 59~days before perihelion and followed
until at least 2023 April~13, 216~days after perihelion (MPC Staff
2022b, 2023).  Near perihelion the comet's brightness peaked at
magnitude~16.  As of the time of this~\mbox{writing}, there has
been no evidence of another outburst.

\begin{figure}[t]
\vspace{0.85cm}
\hspace{-0.85cm}
\centerline{
\scalebox{1.35}{
\includegraphics{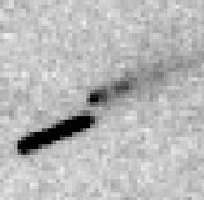}}} 
\vspace{-0.1cm}
\caption{Image of the fragmented comet 157P/Tritton taken by M.\ J\"{a}ger
with his 28-cm f/2.2 reflector at Martinsberg, Austria, on 2022
September 23.12~UT.  Comet 157P-B is 16$^{\prime\prime}$ from the
principal nucleus in a position angle of about 290$^\circ$ and is
aproximately 0.5 mag fainter.  North is up, east to the left.
The field measures some 3$^\prime$ along the diagonal. (Courtesy of
M.\ J\"{a}ger.){\vspace{0.6cm}}}
\end{figure}

\section{Misleading Object 157P-B} 
Discovery of a second condensation was reported by J\"{a}ger
(2022) on images taken with a 28-cm f/2.2 reflector at Martinsberg,
Austria (code G00) on 2022~September 18 and 23 (Figure~4). These and
independent detections of a companion on August~21 through September~2
by the Zwicky Transient Facility (ZTF 2022), using the Oschin 122-cm
f/2.4 Schmidt telescope at Palomar~(code~I41), and on September~15 at
the Xingming Observatory with a 60-cm f/8 Ritchey-Chr\'etien reflector
(code N88) were all published by the MPC at the same time and assigned
to a {\small \bf single object}, designated 157P-B.

\begin{figure}[b]
\vspace{0.97cm}
\hspace{-0.2cm}
\centerline{
\scalebox{0.665}{
\includegraphics{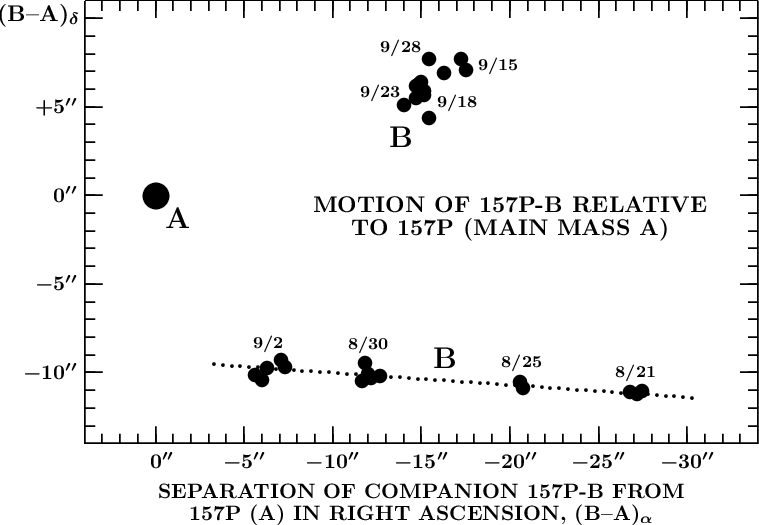}}} 
\vspace{-0.1cm}
\caption{Motion of companion 157P-B relative to the principal comet
157P (marked A) in projection onto the plane of the sky between 2022
August~21 and September~28 (equinox J2000).  The companion was to the
southwest of the main mass in late August and early September, but to
the northwest of it in late September, a dramatic difference.  The
plot shows that there is no way to link all positions by the motion
of a single object.}
\end{figure}

The complete database for 157P-B consists of 30 astrometric observations
on nine nights.  The ZTF contributed 15 data points on four nights, J\"{a}ger
eight points on two nights, and the Xingming Observatory two points on
a single night.  Two observations from the night of September~28 were
subsequently reported from the iTelescope Observatory (code H06).  This
set of 27~observations, including the relevant data for 157P (needed to
compute the separations of 157P-B from 157P in the equatorial coordinates),
has been made available (MPC Staff 2022b).  Three positions for 157P-B
were reported from the SATINO Remote Observatory, Haute Provence (code C95)
on September~26, but because the relevant positions of 157P are missing,
these data could not be used in this investigation.

Figure~5 is a plot of the motion of 157P-B relative~to 157P in right ascension
and declination.  It shows that~an essentially linear trajectory of 157P-B
to the southwest of 157P between August~21 and September~2 was followed
by a nearly stationary trajectory to the northwest in the second half of
September.  Although variable rates of companions' apparent motions are
fairly common among fragmented comets, primarily because of projection
effects, {\small \bf I have never seen a case like this one}.  The perplexing
relative motion of 157P-B in Figure~5 looks like {\small \bf yet another
anomaly} displayed by this comet.

\begin{table}[t]
\vspace{0.2cm}
\hspace{-0.2cm}
\centerline{
\scalebox{0.98}{
\includegraphics{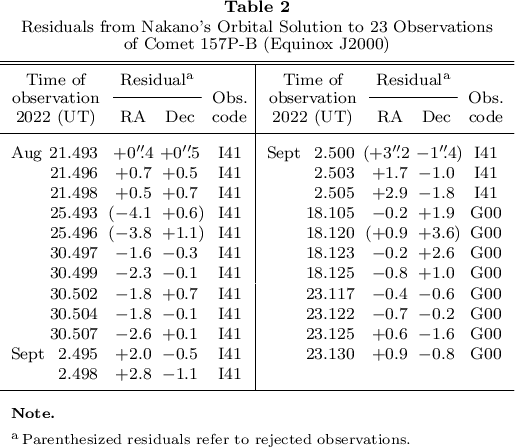}}} 
\vspace{0.6cm}
\end{table}

In this context I should remark that the MPC~Staff (see MPEC 2022-T23
or MPC 160314), the{\vspace{-0.04cm}} JPL Solar System Dynamics
Group,\footnote{See a website
{\tt https://ssd.jpl.nasa.gov/tools/sbdb\_lookup.\\html\#/?des=157P-B},}
and S. Nakano, an associate of the IAU Central Bureau for Astronomical
Telegrams (see footnote 4 below), have independently derived similar
orbits for 157P-B, based primarily or entirely on the astrometry
provided by the ZTF and J\"{a}ger.  Copied in Table~2 are the
residuals left by an orbit computed by Nakano\footnote{The residuals
are copied from J\"{a}ger's message \#30895 to the comets mailing
list, dated 2022 September 25, in which Nakano's orbit for 157P-B
was communicated; for URL see footnote~1.  The file does not appear
to exist among {\it Nakano Notes (NK)\/}.}
from 19~observations between August~21 and September~23.  In fact,
Nakano had 23~observations available but, as the table shows, he rejected
{\it four\/} that left residuals greater than 3$^{\prime\prime}$,
including three observations made with the Palomar Schmidt!  Several
further observations left residuals greater than 2$^{\prime\prime}$
and the mean residual came out to be as high as $\pm$1$^{\prime\prime}\!$.4.
If he did not reject the four observations, the mean residual would have
climbed to $\pm$1$^{\prime\prime}\!$.7.

Not only that residuals of several arcsec left by such~a large fraction
of quality observations have been unheard of in the 21st century, but Table~2
shows a still another peculiarity:\ on a given date, {\it all\/} residuals
tend to be either acceptable (such as August 21 or September 23) or
unacceptable (August 25, August 30 or September 2).  This is in line
with Figure~5, which shows that the same-day relative positions are
consistent, often within 1$^{\prime\prime}$, particularly from the
Palomar Schmidt images.

The arguments of either kind, whether based on evidence from Figure~5
or Table~2, are very compelling,~and when combined they demonstrate
conclusively that the August and late September sets of astrometry
are utterly incompatible and should never have been mixed,~as they
portray the {\small \bf motions of two different companions!}  While
it is not unusual for a split comet to display two or more companions
at the same time, I have no recollection of a previous instance, in
which the variations in activity of two companions of a comet were
synchronized, so that the {\small \bf disappearance of one} was followed
two weeks later by {the \small \bf transient appearance of the other}.
Not only was this exactly what happened with P/Tritton in August--September
2022, but --- as if not enough --- this incredible coincidence fooled
three highly reputable authorities on comet dynamics into computing
orbits of the fake object 157P-B by failing to see through the trick!

\section{Solution to the Problem} 
The verdict is obvious:\ the companion of 157P/Tritton, which was designated
157P-B by the MPC, does not~exist.  Instead, there are two companions:\
one under observation only by the ZTF from August~21 through September~2,
which below is being called {\small \bf companion~C}; and~an\-other,
whose discovery was reported to the MPC by J\"{a}ger and which was
detected between September~15 and 28 also at the Xingming Observatory,
the SATINO Remote Observatory, and the iTelescope Observatory.  In the
following it is referred to as {\small \bf companion~D}.

\begin{table}[b]
\vspace{0.9cm}
\hspace{-0.2cm}
\centerline{
\scalebox{1}{
\includegraphics{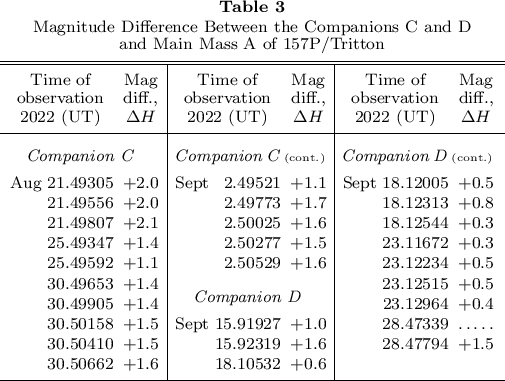}}} 
\vspace{0cm}
\end{table}

Before I employ the software package of my standard model for the split
comets (Sekanina 1978, 1982) to investigate the fragmentation
conditions for companions C and D, I present in Table~3 the measured
magnitude differences, $\Delta H$, between the two companions and the
principal nucleus, based on the observers' reports and listed by the
MPC Staff (2022b):
\begin{equation}
\Delta H = H_{\rm companion} \!-\! H_{\rm principal}.
\end{equation}
Averaging the tabulated brightness measurements, one finds that
\mbox{$\langle \Delta H \rangle = +1.57 \pm 0.29$} mag for companion~C
and \mbox{$\langle \Delta H \rangle = +0.73 \pm 0.46$} mag for D.
Because the brightness is not a robust measure of the companion's
dimensions, the systematic difference does not necessarily suggest
that D is a more sizable fragment than C.

\begin{table*}[t]
\vspace{0.2cm}
\hspace{-0.2cm}
\centerline{
\scalebox{1}{
\includegraphics{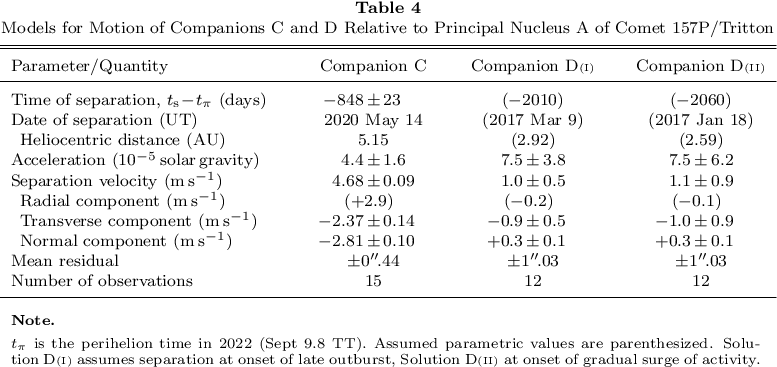}}} 
\vspace{0.75cm}
\end{table*}

\subsection{The Modeling} 
The fragmentation model, allowing one to determine up to five parameters
to fit the motion of a secondary relative to the primary in projection
onto the plane of the sky, was applied separately to companions~C and D.
The short arcs of their appearance available suggest that over most of
their lifetimes both companions were too faint to detect with the
telescopes typically used by the astrometric observers.  In addition,
neither companion was seen to be clearly receding from the main mass; D
was essentially stationary, while C was unquestionably {\it approaching\/}
the main mass in projection onto the plane of the sky.  This kind of
behavior is indicative of fragments that had separated from the parent
nucleus long before (say, on the order of a year or longer) and
were receding (in space) under a low and probably intermittent
outgassing-driven nongravitational acceleration.

Because of the short extent of the orbital arcs over which the companions
were temporarily activated (and therefore observed), it is not expected
that a unique fragmentation solution could be derived for either object.
Rather, I will aim to explore and try to exploit the options that the
fragmentation model offers as well as to employ known properties of the
split comets that past investigations succeeded to establish, in an
effort to {\it constrain\/} the range of credible solutions as much as
possible.  The primary objective is to demonstrate that the observed
motion of either companion could readily be fitted by the fragmentation
model to the degree that a resulting solution offers a distribution of
residuals that is entirely satisfactory, contrary to the unacceptable
residuals in Table~2.  Achievement of this goal will provide the
ultimate proof that the problem of fragmentation of comet 157P/Tritton
has successfully been resolved.

\subsection{Companion C} 
One cannot expect to get a high-quality fragmentation solution from a
12-day arc of the data on companion~C, regardless of their astrometric
quality.  Indeed, a preliminary, reconnaissance run of the model
demonstrated that the radial component, $V_{\rm R}$, of the separation
velocity --- one of the model's five parameters --- was poorly determined.
I used a standard procedure by fitting the trajectory of C for a number
of fixed values of $V_{\rm R}$, an approach that the model allows, and
determined the variations of the other parameters and diagnostic measures
of the quality of fit as a function of $V_{\rm R}$.

The fundamental parameter, the differential nongravitational
acceleration $\gamma$ of the companion, assumed to vary as an
inverse square of heliocentric distance, is always positive (i.e.,
pointing away from the Sun).  Expressed in 10$^{-5}$ units of
the solar gravitational acceleration, it was found to be subjected
to variations with $V_{\rm R}$ (in m~s$^{-1}$) that were with high
accuracy approximated by a polynomial
\begin{equation}
\gamma = 70.00 - 22.71 \:\! V_{\rm R} + 0.083 \, V_{\rm R}^2,
\end{equation}
where the errors of the coefficients were negligibly small.  Positive
accelerations require \mbox{$V_{\rm R} < +3.1$ m s$^{-1}$}.

On{\vspace{-0.06cm}} the other hand, a complete velocity of separation,
\mbox{$V_{\rm sep} = \sqrt{V_{\rm R}^2 \!+\! V_{\rm T}^2 \!+\! V_{\rm N}^2}$}
(where $V_{\rm T}$ is its transverse component and $V_{\rm N}$ its
normal, out-of-plane component),~was found to vary as
\begin{equation}
V_{\rm sep} = 6.63 - 1.18 \, V_{\rm R} + 0.1226 \,V_{\rm R}^2 +
 0.01781 \:\! V_{\rm R}^3.
\end{equation}
The velocity $V_{\rm sep}$ of preferable solutions should be as low
as possible, so I searched for a minimum of this expression. It took
place at \mbox{$V_{\rm R} \!=\! +2.935$ m s$^{-1}\!$} and implied the
values of \mbox{$V_{\rm sep} = 4.67$ m s$^{-1}$} and \mbox{$\gamma =
3.63 \times \! 10^{-5}$\,units} of the solar gravitational acceleration.

On the other hand, the mean residual, $\Re$, was increasing toward
the negative values of $V_{\rm R}$, but at an exceptionally slow
rate,
\begin{equation}
\Re = 0^{\prime\prime}\!.4399 - 0^{\prime\prime}\!.00112 \, V_{\rm R}.
\end{equation}
More significant was the magnitude of the mean residual --- the
Palomar observations of the relative motion of companion~C were
fitted with a mean residual of $\pm$0$^{\prime\prime\!}$.44 in a
wide range of $V_{\rm R}$, confirming their high accuracy.

All solutions for \mbox{$V_{\rm R} \sim +2.9$ m s$^{-1}$} left
essentially identical residuals.  This value of $V_{\rm R}$ was
selected below for the representative solution.  Its interesting
feature is that companion~C separated from the parent nucleus
around mid-May 2020, about three months after the comet had
undergone a close encounter with Jupiter, with a miss distance
of only 0.265~AU.

\begin{table*}[t]
\vspace{0.2cm}
\hspace{-0.2cm}
\centerline{
\scalebox{1}{
\includegraphics{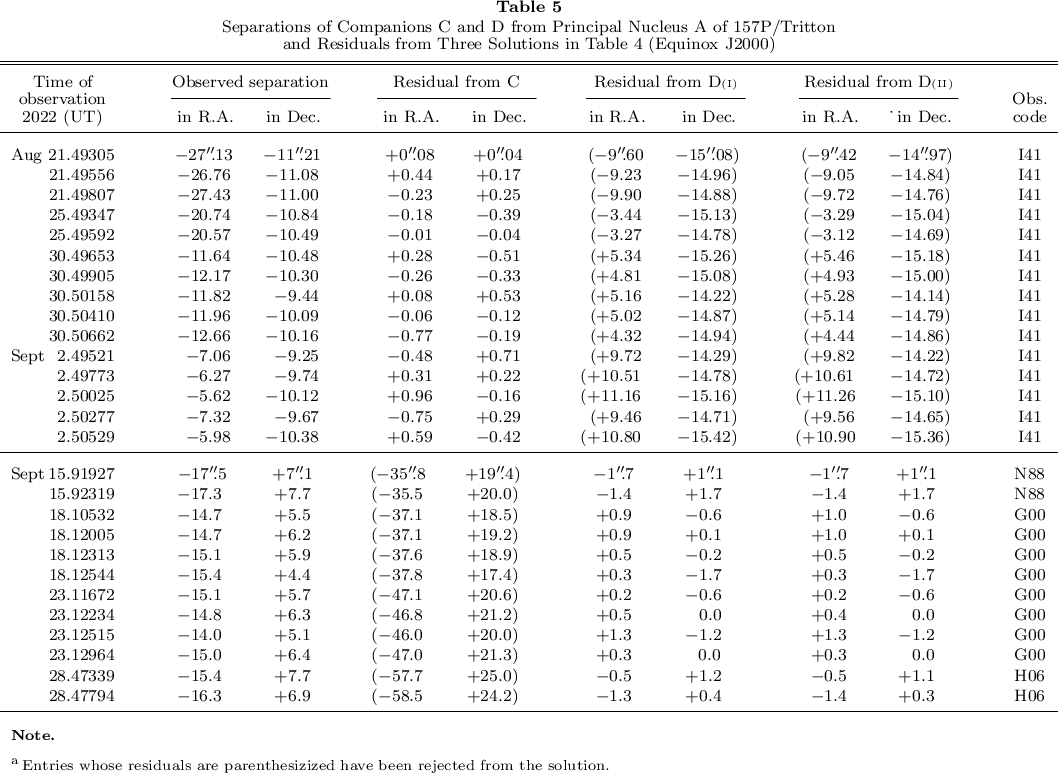}}} 
\vspace{0.68cm}
\end{table*}\
The parameters of the adopted fragmentation solution for companion~C
are presented in Table~4, while the residuals from the Palomar
observations are listed in columns 4--5 of Table~5, which also shows
the  major deviation of the motion of C from the trajectory of the
other companion.  By the time of the last observation, the position
was off by more than 1$^\prime$.  Note that the nongravitational
acceleration is predicted to be low, meaning the companion may
survive until the next return to perihelion.

An ephemeris for companion C suggests that its projected distance
from the principal mass reached a peak of nearly 100$^{\prime\prime}$
in position angle 205$^{\circ}$ at the beginning of
September 2021, when the comet was under observation, even though
it was extremely faint (mag $\sim$21).  Another peak of 108$^{\prime\prime}$
in position angle 226$^{\circ}$ occurred in early May 2022,
when the comet could not be observed.  Interestingly, the minimum
separation, 10$^{\prime\prime}$ to the south of the principal nucleus,
took place only a few days after the last reported observation at
Palomar.

\subsection{Companion D} 
A look at Figure 5 makes it clear that the fragmentation solution
for companion D, to the northwest of the principal mass (A), should
still be much more uncertain than the solution for companion~C.
Orientation runs that included the radial ($V_{\rm R}$) and/or
transverse ($V_{\rm T}$) components of the separation velocity
as variables did indeed fail to provide meaningful parameters for
companion~D.  Under these circumstances, the starting working run
was one in which I solved for the separation time, the acceleration,
and the normal component of the separation velocity.  The acceleration
came out to be lower than 10$^{-5}$\,unit of the solar gravitational
acceleration (probably unrealistically low) and{\vspace{-0.03cm}} the
out-of-plane separation velocity was lower than 1~m~s$^{-1}$.  The most
surprising~was the resulting separation time:\ although determined with
a large uncertainty, the nominal date was 2017 January~6, consistent with
the constraints on the timing of a post-perihelion brightening in early
2017, as established in Section~2.  It appears that just like in many
other comets, the brightening and the breakup were correlated.

If so, the companion was likely to have separated from the parent
nucleus at the onset of brightening, which~may have been either (i)~a
gradual surge of activity with progressively increasing amount of
material in the coma (as displayed by comet C/2019~Y4), in which
case the onset was probably nearer the beginning of the 80-day long
period of uncertainty, i.e., in January 2017; or it was (ii)~a
typical outburst, with an extremely rapid raise of activity, in
which case the onset time could be at any point of the period of
uncertainty, including just a few days before 19~March 2017.

Accordingly, I chose to run two models for companion~D with very
different times of separation:\ one --- D{\tiny (I)} --- on 2017 March~9,
or 2010~days before the 2022 perihelion time; the other --- D{\tiny
(II)} --- 50~days earlier, on January~18, or 2060~days before perihelion;
looking for a potential effect on the companion's motion.  Such runs
were still examined as a function of the radial component, $V_{\rm R}$,
of the separation velocity, but the three remaining parameters,
$\gamma$, $V_{\rm T}$, and $V_{\rm N}$, could have been solved for.

The nongravitational acceleration varied rapidly, but essentially
linearly, with $V_{\rm R}$.  In the same units as before,
\begin{eqnarray}
\gamma & = & 2.83 - 23.14 \, V_{\rm R} \;\;\; {\rm (for} \;
  t_{\rm s} \!-\! t_\pi = -2010 \; {\rm days)} \nonumber \\[0.05cm]
\gamma & = & 4.59 - 29.49 \, V_{\rm R} \;\;\; {\rm (for} \;
  t_{\rm s} \!-\! t_\pi = -2060 \; {\rm days),}
\end{eqnarray}
where $t_{\rm s}$ is the separation time of companion~D and $t_\pi$
is the comet's perihelion time in 2022.  Similarly, the mean residual
varied as
\begin{eqnarray}
\Re & = & 1^{\prime\prime}\!.0265 - 0^{\prime\prime}\!.0055 \, V_{\rm R}
 \;\;\; {\rm (for} \; t_{\rm s} \!-\! t_\pi = -2010 \; {\rm days)}
 \nonumber \\[0.05cm]
\Re & = & 1^{\prime\prime}\!.0260 - 0^{\prime\prime}\!.0075 \, V_{\rm R}
 \;\;\; {\rm (for} \; t_{\rm s} \!-\! t_\pi = -2060 \; {\rm days).}
 \nonumber \\[-0.1cm]
 & &
\end{eqnarray}
Both $V_{\rm T}$ and $V_{\rm N}$ came out to be very low, not exceeding
1~m~s$^{-1}$, so the separation velocity was not an issue.  Surviving
longer than 5.5~years, D was a persistent companion (see Sekanina
1982) and its acceleration $\gamma$ could hardly exceed $\sim \! 10 \times
\! 10^{-5}$\,units of the solar gravitational acceleration.  Adopting
for either model a $\gamma$ value of 7.5, the radial component of the
separation velocity stays in the range of 0.1--0.2~m~s$^{-1}$.

The parameters of the two fragmentation models presented in Table~4
show that the only major difference~is the much higher errors in the
case of Solution D{\tiny (II)}.  However, in either case the parametric
accuracy is unsatisfactory, which is necessarily the result of a very
short period of time covered by the observations.  Table~5 demonstrates
that the distributions of residuals from the two solutions are practically
identical.  Given that most astrometric observations were made with
a telescope whose resolving power was more than 300$^{\prime\prime}$
per mm, the residuals look more than satisfactory.  There are possibly
slight, 1$^{\prime\prime}$--2$^{\prime\prime}$, systematic differences
between the Martinsberg astrometry on the one hand and the Xingming
and Mayhill astrometry on the other hand, but this should not be all
that surprising, considering that the companion was a difficult object
to measure (Figure~4), especially in telescopes of smaller sizes.

The uncertainties of the derived fragmentation solutions notwithstanding,
the problem of 157P-B has clearly been successfully resolved.

\section{Conclusions} 
The unorthodox behavior of comet 157P/Tritton culminated in 2022,
when each of two companions that had separated at vastly different
times brightened enough over short periods of time to be observable
with telescopes of modest sizes.  Two weeks after one of the
companions faded, the other brightened for two weeks only to
fade as well.  The perfect coordination of the two entirely
independent events made a false impression as if a single
object, designated 157P-B by the MPC, was intermittently visible.
The fundamental differences, such as the positions and motions of
the two companions relative to the principal mass, which exhibited
unmistakable signs of incompatibility, were overlooked or ignored.

Among the results of the mixup were the meaningless orbits computed
by highly reputable authorities from the positions of the two objects.
As expected, the orbits left large residuals (up to 4$^{\prime\prime}$)
from the astrometric positions that were accurate to a fraction
of 1$^{\prime\prime}$.

Because of the short lengths of the observed orbital arcs of the
two companions, their fragmentation parameters could not accurately
be derived, only constrained.  Yet, it is likely that one of the
two objects separated from the parent nucleus at the time of the
major outburst or surge of activity in the first quarter of 2017.
The other companion appears to have detached in 2020, following
the comet's close encounter with Jupiter.  Both objects appear to
be persistent companions with low nongravitational accelerations.

This investigation has touched upon two issues worth commenting
on.  One is emphasis on negative observations of faint comets,
which should be (but are seldom) reported.  When a comet is caught
in outburst or surge of activity, negative observations preceding
the event can prove extremely useful in constraining its onset time.

The other issue is the idea of launching a~\mbox{campaign}~of imaging
comets, with a powerful instrument (such~as~the HST's wide-field
camera), at nearly random~times,~but at least months after a
major outburst;~this~could~prove highly rewarding in terms of
detecting sizable~but~excep\-tionally faint fragments of the
nucleus~(apparent~mag\-ni\-tude $\gg$21) that otherwise remain undetected.\\

\begin{center}
{\footnotesize REFERENCES}
\end{center}
\vspace{-0.4cm}
\begin{description}
{\footnotesize
\item[\hspace{-0.3cm}]
Green, D.\ W.\ E.\ 2003, IAU Circ. 8215
\\[-0.57cm]
\item[\hspace{-0.3cm}]
J\"{a}ger, M. 2022, Minor Plan.\ Electr.\ Circ.\ 2022-T23
\\[-0.57cm]
\item[\hspace{-0.3cm}]
Marsden, B.\ G.\ 1979, Catalogue of Cometary Orbits, 3rd ed.{\linebreak}
 {\hspace*{-0.6cm}}Cambridge, MA:\ Smithsonian Astrophysical Observatory,
 88\,pp
\\[-0.57cm]
\item[\hspace{-0.3cm}]
Marsden, B.\ G.\ 1995, Catalogue of Cometary Orbits, 10th ed.{\linebreak}
 {\hspace*{-0.6cm}}Cambridge, MA:\ IAU Central Bureau for Astronomical
 Tele-{\linebreak}
 {\hspace*{-0.6cm}}grams \& Minor Planet Center, 108\,pp
\\[-0.57cm]
\item[\hspace{-0.3cm}]
Marsden,\,B.\,G.,\,\&\,Green,\,D.\,W.\,E.\,1985,
 Quart.\,J.\,Roy.\,Astron.\,Soc.,{\linebreak}
 {\hspace*{-0.6cm}}26, 92
\\[-0.57cm]
%
%
%
\item[\hspace{-0.3cm}]
Minor Planet Center Staff 2017, Minor Plan.\,Circ.\,104100,\,104101,{\linebreak}
 {\hspace*{-0.6cm}}104983, 105338, and 105699
\\[-0.57cm]
\item[\hspace{-0.3cm}]
Minor Planet Center Staff 2021, Minor Plan.\,Circ.\,132831,\,134067,{\linebreak}
 {\hspace*{-0.6cm}}and 135329
\\[-0.57cm]
\item[\hspace{-0.3cm}]
Minor Planet Center Staff 2022a, Minor Plan.\ Circ.\ 136702 and{\linebreak}
 {\hspace*{-0.6cm}}142106
\\[-0.57cm]
\item[\hspace{-0.3cm}]
Minor Planet Center Staff 2022b, Minor Plan.\,Circ.\,141285,\,142106,{\linebreak}
 {\hspace*{-0.6cm}}158657, and 158658
\\[-0.57cm]
\item[\hspace{-0.3cm}]
Minor Planet Center Staff 2023, Minor Plan.\ Circ.\ 160514,\,162180,{\linebreak}
 {\hspace*{-0.6cm}}and 163393
\\[-0.57cm]
%
%
\item[\hspace{-0.3cm}]
Sekanina, Z.\ 1978, Icarus, 33, 173
\\[-0.57cm]
\item[\hspace{-0.3cm}]
Sekanina, Z.\ 1982, in Comets, ed.\ L.\ L.\ Wilkening (Tucson:\ Univer-{\linebreak}
 {\hspace*{-0.6cm}}sity of Arizona Press), 251
\\[-0.64cm]
%
%
\item[\hspace{-0.3cm}]
Zwicky Transient Facility 2022, Minor Plan.\ Electr.\ Circ.\ 2022-T23}
\vspace{-0.745cm}
\end{description}
\end{document}